# Electron-phonon interaction and point contact enhanced superconductivity in trigonal PtBi$_2$.

D. L. Bashlakov[1], O.E. Kvitnitskaya[1], G. Shipunov[2],
S. Aswartham[2], O.D.Feya [2,3], D.V. Efremov[2], B. Büchner[2,4], Yu.G. Naidyuk[1].

[1]B. Verkin Institute for Low Temperature Physics and Engineering, NAS of Ukraine, 61103 Kharkiv, Ukraine

[2]Institute for Solid State Research, IFW Dresden, D-01171 Dresden, Germany

[3] Kyiv Academic University, 03142, Kyiv, Ukraine

[4]Institut für Festkörper- und Materialphysik and Würzburg-Dresden Cluster of Excellence ct.qmat, Technische Universität Dresden, 01062 Dresden, Germany

**Abstract**

PtBi$_2$ is a Weyl semimetal, which demonstrates superconductivity with low critical temperature $T_c$ ~ 0.6 K in the bulk. Here, we report our study of electron-phonon interaction (EPI) in trigonal PtBi$_2$ by the Yanson point contact (PC) spectroscopy and presenting the observation of PC enhanced superconductivity. We show, that the Yansons PC spectra display a broad maximum around 15 meV, indicating, apparently, EPI mechanism of Cooper pairing in PtBi$_2$. Moreover, we discovered a substantial increase of $T_c$ up to ~ 3.5 K in PCs. The observed $T_c$ is sufficiently higher than the bulk value, as well as  detected at hydrostatic pressure. We calculated the phonon density of states and Eliashberg EPI function in PtBi$_2$ within the framework of the density functional theory. A comparison of experimental data with theoretical calculations showed acceptable agreement. The theoretical $T_c$ is 3.5 K, which corresponds to the experimental value.

**Introduction**

Currently, solid state physics is facing a growing demand for new quantum materials with technical prospects. Emerging topological superconductors are among such new quantum materials, which could find applications in next-generation devices for fault-tolerant quantum computing and/or quantum cryptography [1-3]. There are a number of topological Weyl semimetals that exhibit superconductivity at low temperatures. Because of the inversion symmetry breaking, superconductivity in materials of this type is always considered a candidate for topological superconductivity [3].

$PtBi_2$ is one of such candidates showing the traces of superconductivity [4] as well as presence of Dirac-like states [5-8] and triply degenerate Weyl points near the Fermi level [8]. Semimetallic $PtBi_2$ is known to exist in polymorphic forms of cubic and trigonal structures [4]. Giant positive magneto-resistance is a commonly shared property for both of them [9-11]. Similarly, both phases demonstrate the transition in to superconducting (SC) state with $T_c$ rising by applying pressure [12, 13]. While the cubic phase is substantially a 3D material, the trigonal one is a layered material. This layered nature of quasi two-dimensional crystal structure of the $PtBi_2$ gives advantages for device fabrication. Also, trigonal polymorph of $PtBi_2$ can be easily sized down to the few layers just by mechanical exfoliation technique. The parent compound of trigonal $PtBi_2$ is SC at $T_c \sim 0.6$ K. It was also shown, that doping of the trigonal phase with rhodium (Rh) causes the increase of $T_c$ up to 2.5 K [4]. The cyclotron mass for $PtBi_2$ was shown to be about 0.13 $m_e$ confirming a high mobility of the charge carriers [14]. Also, the $PtBi_2$ is an ideal platform to study 2D-dimensional superconductivity in a topological semimetal [14]. However, the mechanism of superconductivity in layered $PtBi_2$ is still unclear. Here, for the first time, we have measured the electron-phonon interaction (EPI) in $PtBi_2$ using Yanson point-contact (PC) spectroscopy [15]. We supplemented the experimental study with a theoretical calculation of EPI function.

Previous, PC studies on a number of topological materials demonstrated local enhancement of SC transition temperature [16-27]. In this work we are adding $PtBi_2$ into the list of topological materials which show local enhancement of $T_c$ in the PC with normal metals. The temperature of transition in the SC state in the hetero-contacts between $PtBi_2$ and of Au, Ag or Cu rises up to 3.5 K and critical magnetic field of up to 3 Tesla (see, e.g., Fig.1S in Supplement [28]) is needed to suppress the observed superconductivity.

**Experimental and results**

Samples of trigonal PtBi$_2$ were grown via self-flux method. Single crystals of PtBi$_2$ were synthesized by mixing elemental powders of Pt (99.99%, Saxonia Edelmetalle GmbH), and Bi (325-mesh powder, 99.5%, Alfa Aesar). More details on the crystal growth and characterization can be found in [4]. Details of PC preparation and measurement technique are given in "Supplement" [28].

Yansons PC spectra (second derivative of current-voltage characteristic [15]) measured for the PtBi$_2$ crystals with residual-resistivity ratio RRR≈130 are shown on Fig.1. Spectra were obtained for homo-contacts (grey lines) and hetero-contacts (black lines) with gold or silver thin wires (counterelectrodes). Yansons PC spectra of homo-contacts (grey lines) demonstrate some smeared humps or kinks in the region of 10-30 mV, while hetero-contacts (black lines) developed distinctly pronounced maximum.

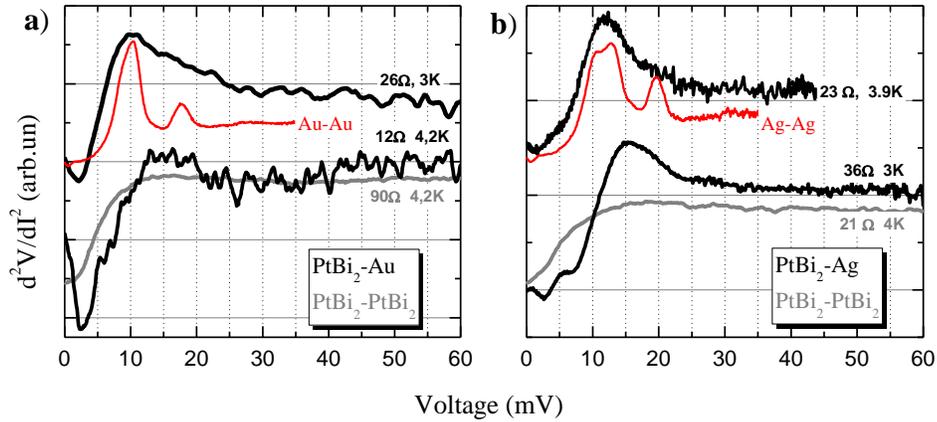

**Figure 1**. Yansons PC spectra ($d^2V/dI^2$) of the hetero-contacts PtBi$_2$ (black lines) made with gold **a)** and silver **b)** counter-electrodes. Spectra of PtBi$_2$ homo-contacts (grey lines) on both graphs are shown for comparison. Blue and red curves are Yanson PC EPI spectra for Au and Ag from [29].

From one of the pioneering work [30] it is known that phonon peaks of both materials constituting PC are present on Yansons PC spectrum. Indeed, one can see that some selected spectra (top ones from Fig. 1a and 1b) have a dominated input where the transverse phonon maximum of gold or silver is expected [29-32]. At the same time, the bottom spectra of PtBi$_2$ hetero-contacts shown here demonstrate well defined maximum around 15 mV, which can not be assigned to silver or gold. It is clear that such spectral feature correlates with the hump noted for the PtBi$_2$-PtBi$_2$ contacts. Thus, we can relate it to the EPI in PtBi$_2$.

Most of the Yansons PC spectra were collected at temperatures near to the helium boiling point. A majority of the PCs demonstrate transition to the SC state at 3–3.5 K. Fig. 2 (a,b) shows examples of the differential resistance $R_d(V)=dV/dI$ with zero-bias dips for some selected hetero-contacts, which disappeared with increasing of temperature or magnetic field (see Fig. S2 [28]) We observed such behavior both for hard and soft PCs, which were made with conducting silver paint[*]. The distribution of temperatures and magnetic field at which the SC dips in the $dV/dI$ were suppressed is shown on Fig. 2c.

**Discussion**

The PC spectra of $PtBi_2$ homo-contacts have a broad maximum and a high background level at the energy above the phonon. This suggests that the material inside the PC may be strained, probably due to deformations obtained during mechanical PC creation. The use of softer materials, such as noble metals, improved the resolution of the spectral features.

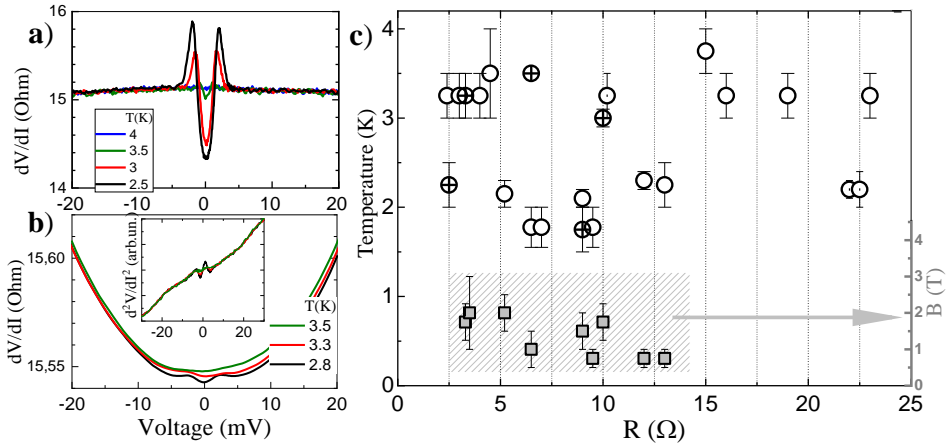

**Figure 2.** $dV/dI$ of $PtBi_2$–Ag hard **a)** and soft **b)** PCs . The $dV/dI$ data for the Fig. b) were recalculated from the corresponding second derivative $d^2V/dI^2$ in inset. **c)** SC transition temperature (linked to the left scale) detected within various hard PCs between $PtBi_2$ and Ag (open circles), Au (crossed circles) or Cu (circles with a dot) counter-electrodes, as well as measured at 1.55 K in a magnetic field (square symbols linked to the right scale) needed to suppress superconductivity in PCs.

---

[*] More details on how contacts were made and some more examples of the $dV/dI$ with SC zero-bias dips as well as more Yansons PC spectra of $PtBi_2$-Me (where Me –Au, Ag or Cu) heterocontacts and $PtBi_2$-$PtBi_2$ homocontacts are given in Figures at "Supplement" [28].

Another effect to note is a varying partial contribution to the Yansons PC spectrum from each material in the heterocontacts. The contribution to the spectra from each material is determined by the geometry of the contact (a fraction of the volume occupied by each metal in the PC core). Therefore, the intensity of phonon peaks of each material is, as a rule[†], proportional to the effective volume occupied by each metal [32, 33]. Here, for hetero-contacts, the Au (Ag or Cu) tips can make a contribution to the Yansons PC spectra. For these reasons, the data from the $PtBi_2$–Cu hetero-contacts were not included, because the major EPI peaks of Cu are at 15–20 meV [30, 32], which could mask the contribution from $PtBi_2$.

Figure 3 demonstrates calculated phonon density of states (DOS) and the Eliashberg EPI function $\alpha^2F(\omega)$ of $PtBi_2$. To calculate phonon we used density functional theory (DFT) [34,35] implemented in the VASP code [36–38] with PHONOPY package [39]. Perdew-Burke-Ernzerhof (PBE) generalized gradient approximation (GGA) [40] and projector-augmented wave (PAW) [41,42] within DFT was used for these calculations. We have set the wave kinetic energy cutoff to 300 eV, and uniform Γ-centered 4x4x4 k-mesh which were used for Brillouin zone sampling. Structural relaxation was stopped with forces acting on ions achieved $10^{-5}$ eV/Å accuracy. To calculate superconducting properties of $PtBi_2$ we used DFT in conjunction with density functional perturbation theory implemented in Quantum Espresso code [43,44]. The valence electronic wave functions were expanded with optimized kinetic-energy cutoff 110 Ry. The core-valence interaction is taken into account with scalar relativistic ultrasoft GGA PAW pseudopotentials. We employed gaussian Fermi surface smearing of 0.01 Ry in the Brillouin zone integration. Optimized structural parameters were a=12.6428 Bohr (6.6903 Å) and c/a=0.9439, which leads to c=11.9335 Bohr (6.3148 Å). We used 8x8x8 k-points mesh for structural relaxations and charge density calculations and 4x4x4 q-grid for phonon calculations within density functional perturbation theory with interpolated phonon-electron calculation scheme. The superconducting transition temperature ($T_c$), EPI strength (λ) and Eliashber EPI function $\alpha^2F(\omega)$ was obtained with Migdal-Eliashberg theory [45] and McMillan-Allen-Dynes empirical formula to estimate $T_c$ [46,47]. The calculated $T_c$ from EPI is 3.5 K for the screened-repulsive interaction $\mu^*=0$ and $T_c=0.4$K for $\mu^*=0.18$.

---

[†] Partial intensity of EPI features for each metal in Yansons PC spectrum of heterocontact depends also from intensity of EPI in specific material and is inversely proportional to the Fermi velocity [15]. Because of the EPI strength and the Fermi parameters in $PtBi_2$ are less known, it is very speculative to predict which EPI features will prevail in the case of "geometrically symmetrical" hetero contact.

It is seen that the maximal energy of phonons is a bit above 18 meV and above 12 meV three bands of optical phonons at around 13, 15 and 18 meV are seen. Smoothing of the phonon DOS results in appearance of broader maximum at around 15 meV. EPI function $\alpha^2F(\omega)$ looks similar to the phonon DOS. Comparing $\alpha^2F(\omega)$ with measured Yansons PC spectrum in Fig. 4(a), we see that mainly optical phonons contribute to EPI, while low energy phonons below 12 meV are not resolved in the PC spectra. There are also some shallow features on Yansons PC spectrum as a little maximum at 5 mV and shoulder at 25-30 mV. However, these features are not reproduced for all other spectra shown in Fig. 1 and Figs. S2, S3 in Supplement. Therefore, these features can be some artefact. Although shoulder at 25-30 mV maybe due two-phonon processes [15], which reflect the main maximum at 15 mV.

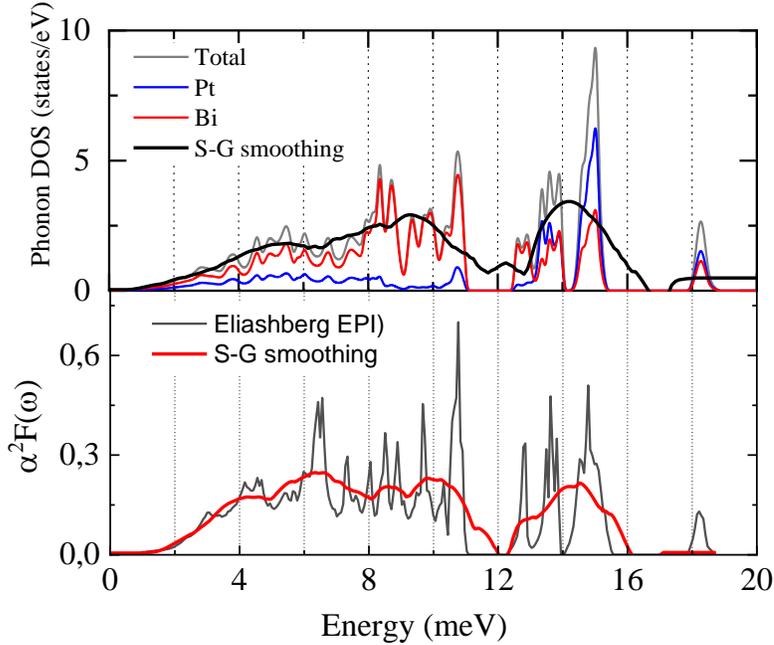

**Figure 3**. Top panel: Calculated phonon DOS of PtBi$_2$ (grey line) with partial contribution from Pt (blue) and Bi (red) phonons. Thick (black) curve is smoothed with Savitzky-Golay digital filter phonon DOS of PtBi$_2$. Bottom panel: The calculated and smoothed Eliasherg EPI function $\alpha^2F(\omega)$.

On Fig. 4(b) we compare Yansons PC spectrum of PtBi$_2$ with that of MoTe$_2$. The latter is also a layered Weyl semimetal demonstrating enhanced $T_c$ in PC with normal metals [22, 23]. Sharp peak close to a zero-bias is a rudiment of superconductivity present in PC. Similar peak was observed for PtBi$_2$ PCs cooled below $T_c$ . Comparing the Yansons PC spectra of

PtBi$_2$ and MoTe$_2$ on Fig. 4(b) one can see certain similarity between them, although atomic mass of Pt and Bi is much larger than Mo and Te. Both compounds have the phonons with the definite energy, which contributes the most in to the spectra. We can not state at this stage that it is a common trend for the layered compounds demonstrating PC stimulated/enhanced superconductivity and having a van-der-Waals interlayer coupling. We believe that future studies should also focus on looking for a dominant lattice vibrations, since they can be responsible for the Cooper pairing. Unfortunately, majority measurements of PC with enhanced superconductivity [16-21, 23, 24-27] are focused on differential resistance/conductivity of the PCs ignoring the Yansons PC spectroscopy.

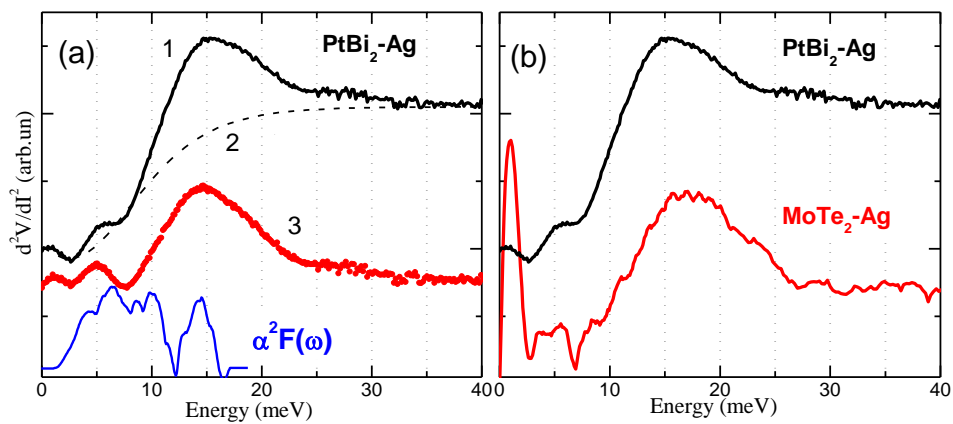

**Figure 4**. (a) Yansons PC spectrum of PtBi$_2$ –Ag heterocontact (1) after background (2) subtraction (3) ($T$=3K, $B$=0T). Bottom curve is smoothed Eliashberg EPI function $\alpha^2F(\omega)$ from Fig.3. (b) Same Yansons PC spectrum in comparison with spectrum of another layered Weyl semimetal MoTe$_2$ [22] ($T$=2.3K, $B$=0.4T). Sharp peak near zero-bias is due to residual superconductivity in PC.

The SC dips in the *dV/dI* of PtBi$_2$ contacts showed no characteristic Andreev reflection structure (see Fig. 2a, 2b and Fig. S1 [28]), by that we mean absence of a double minima structure typical for S-N contacts as in the case of MoTe$_2$ [22]. It is possible speculate that this is related to the temperature smearing of the spectrum [48], to a low barrier value (*Z*) on the N-S interface [48, 49], to an unconventional order parameter [50]. However sharp side peaks with featureless zero-bias dip is characteristic for nonspectral regime in PC when SC state is suppressed by high current density, which increases with voltage and results also in the temperature rise in PC [51]. But certainly PtBi$_2$ belongs to the list of materials with nontrivial topology where PC enhanced/induced superconductivity was detected such as TaAs [16, 17], TaP[18], Au$_2$Pb[25], Cd$_3$As$_2$ [19-21], MoTe$_2$ [22, 23], WTe$_2$ [24], WC [26],

TaAs$_2$ and NbAs$_2$ [27]. Interesting to note that our preliminary studies of PCs with rhodium doped crystals of PtBi$_2$ demonstrated T$_c$ comparable to what was observed in a bulk for this samples (T$_c$ close to 3 K measured for crystals with 35% of platinum substituted by rhodium) [4]. Although the magnetic field needed to suppress this superconductivity at 1.5 K was twice higher from one we observed in the contacts for pure PtBi$_2$ (see Fig. S4 in Supplement [28]).

PC enhanced *T$_c$* was also detected in the iron-based superconductors [52, 53], but that was likely related to the surface "*rather modification then degradation*". In contrast to iron-based superconductors, studied ditellurides are stable at ambient atmosphere [54-56] as well as tungsten carbide [57] due to the passivation layer formed on the surface. Also having tungsten carbide on a list we can rethink on the local pressure effect. Tungsten carbide is one of the hardest materials and is unlikely to be easily compressed by metals like gold [26] enough to make changes in the crystal lattice causing the T$_c$ rise.

It appears that PC itself is not a single way to stimulate SC state in topological materials. Recently it was reported that adding a thin layer of normal metal stimulates the surface superconductivity in some semimetals from the above list, in particular for: WTe$_2$/Pd [58], WC/Me (Me – Au, Pt, Fe, Co, Ni) [59] and TaAs/Ag [60]. It is hard to believe that pressure effect plays role in the increase of the SC temperature for these materials. The changes in the electronic state of semimetals on the interface with the metals more likely is the reason of the observed phenomena. Solid proof for this point is the present result on an encapsulated in the boron nitride semimetallic molybdenum ditelluride, where T$_c$ was raised up to 7.6 K in the device containing a single layer of MoTe$_2$ [61]. In this paper, authors assign electronic interaction in MoTe$_2$ to be responsible for the observed enhanced superconductivity.

**Conclusions**

We studied EPI in semimetallic trigonal PtBi$_2$. Yansons PC spectra reveal presence of the maximum at around 15 meV, which dominates in EPI. Comparison of Yansons PC spectra with calculated EPI function $\alpha^2 F(\omega)$ demonstrates that the high energy phonons mostly contribute to the EPI. No explicit other features were noticed, thus EPI might be responsible for the Cooper pairing in this compound. Induced by PCs superconductivity in trigonal PtBi$_2$ have transition temperature as high as 3.5K, which is few times higher than the bulk value. The observed in PCs T$_c$ also exceeds the previously reported values in case of system been doped PtBi$_2$ with Rh (2.8-3K) [4] or placed under the hydrostatic pressure (~2K) [13]. We speculate that, the observed enhanced $T_c$ realizes due to changes in the LDOS of electrons induced by the normal metal/semimetal interface. Although, we cannot rule out effect of a local pressure created by the contacts, which can also contribute to increase of $T_c$. Our calculated $T_c$ from the EPI is 3.5 K, which agree well with $T_c$ in the PC experiments.

**Acknowledgement**

We are grateful to S. Gaβ and A. Wolter-Giraud for the technical assistance and J. Dufouleur for stimulating discussions. We would like to acknowledge funding by Volkswagen Foundation. DB, YuN and OK are also grateful for support by the National Academy of Sciences of Ukraine under project Φ19-5 and thankful to the IFW Dresden for hospitality. SA appreciating support from Deutsche Forschungsgemeinschaft (DFG) through Grant No: AS 523/4-1.

# SUPPLEMENT
## to the paper by D. L. Bashlakov et al. "Electron-phonon interaction and point contact enhanced superconductivity in trigonal PtBi$_2$"

**Experimental details**

Samples used in experiments where grown in the same batch. Details on the sample synthesis and transport characterization can be found in [4]. Visually both crystals had some remaining flux attached to their sides. Smaller crystal was about dozen of micron thick and been used for collection of the preliminary results – and contacts were made to the side of the sample which had no flux remained. Bigger crystal was "three-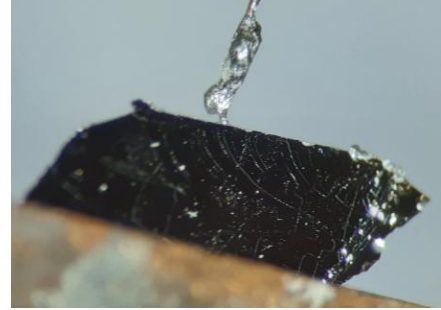dimensional" in a way that its thickness was comparable with in-plain sizes. Thus it was possible to slice it on to thinner pieces and make the contacts to freshly exposed surfaces. Although, we did not see a difference between the data collected for the freshly cleaved surfaces and the flux free ones, as well as no qualitive difference in data was noticed between two crystals.[ADS1]The data were collected on two LHe setups with lowest available temperature of 2.5 K and 1.55 K with magnetic field available of 15 Tesla and 4 Tesla accordingly. Hard contact where made directly in cryostat by mechanically touching the wire of Cu, Ag or Au to the edge or to the plane of a crystal. Soft contacts where made by placing a drop of a conductive Ag paint between the sample and 0.07mm copper wire (see image on the right). Thus soft contact where prepared at ambient atmosphere and temperature and moved in to cryostat only after silver paint was dried out.

We measured the current-voltage (*I–V*) characteristics of PCs and their first *dV/dI* and second $d^2V/dI^2$ derivatives by look-in technique. The *dV/dI(V)*≡$R_d(V)$ and the $d^2V/dI^2(V)$ were recorded by sweeping the dc current *I* on which a small ac current *i* was superimposed.

## Additional experimental results

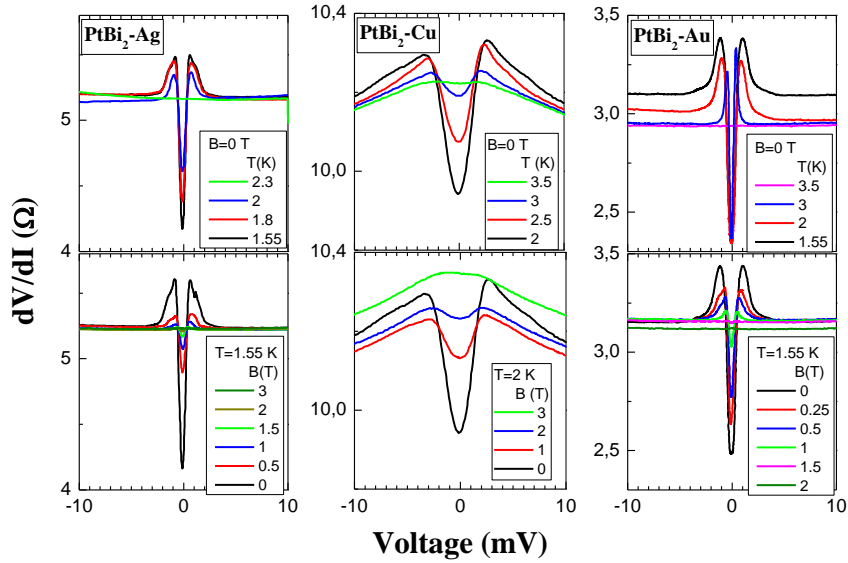

**Figure S1.** Differential resistance from several contacts PtBi$_2$-Me (where Me – Ag, Cu or Au) measured at various temperatures (top) and magnetic fields (bottom).

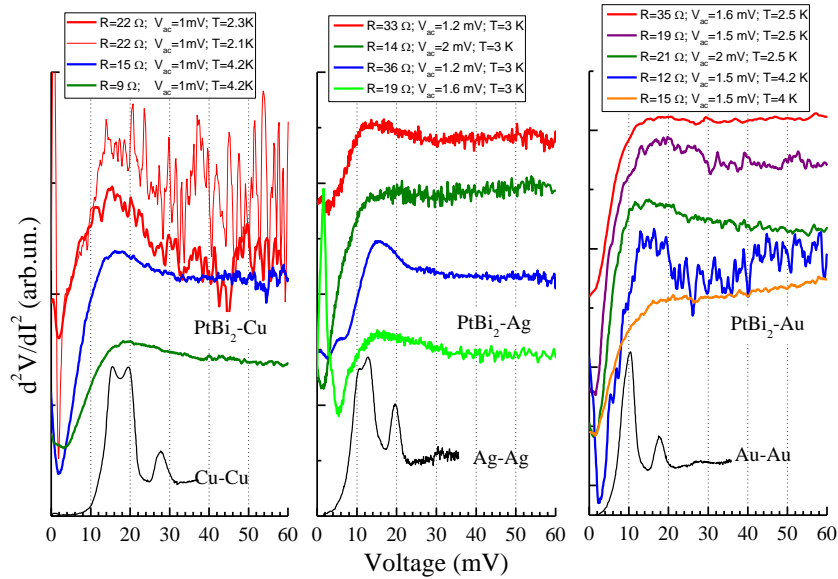

**Figure S2.** Yansons PC spectra of the hetero-contacts, some spectra shifted up or down for a clarity. Left image shows spectra for PtBi$_2$-Cu PC where red curves belong to same contact taken at different temperatures (thing red line has zero bias anomaly due do excess SC current). Middle and right images shows spectra for PtBi$_2$-Ag and PtBi$_2$-Au contacts accordingly. The screening factor for the selection of the data for the last two sets was – spectra should not contain feature that may belong to Ag or Au , i.e. there should not be dominating peeks present at 10-13mV (Ag) and 9-11mV (Au) (Yansons PC spectra for the noble metals shown on the corresponding images with the thin black lines).

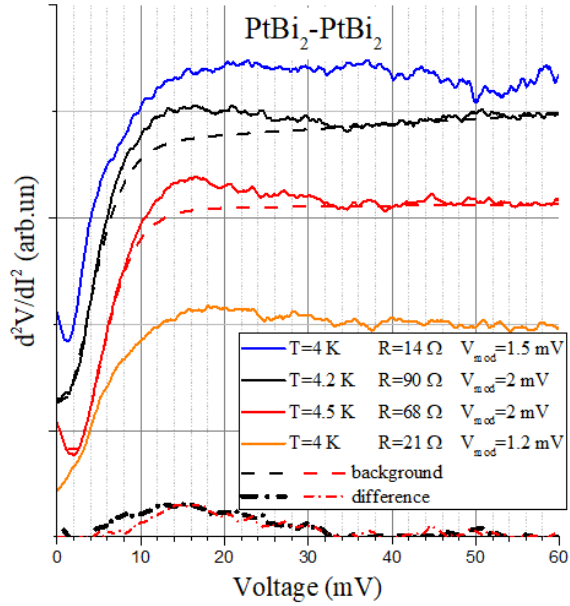

**Figure S3.** Yansons PC spectra of the homocontacts PtBi$_2$-PtBi$_2$, some spectra shifted either up or down for a clarity. The background for the red and black curves was calculated by the formula: $BG = \alpha\, \text{th}^2(\beta(V+\delta)) + \chi V$, where $V$ is voltage and $\alpha$, $\beta$, $\delta$ and $\chi$ are the fitting parameters.

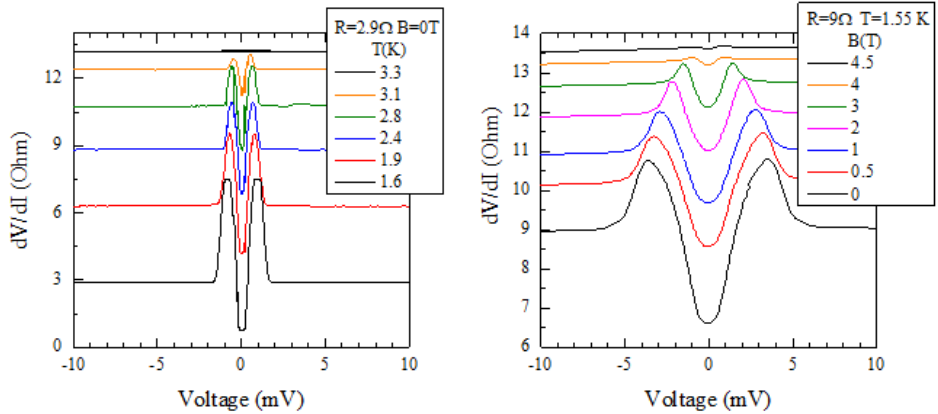

**Figure S4.** Differential resistance of the PCs made between the Pt$_{0.65}$Rh$_{0.35}$Bi$_2$ single crystal and the gold counter-electrode measured at various temperatures (left) and magnetic fields (right).